\documentclass[preprint,aps]{revtex4}

\usepackage{bm}

\begin{document}


\title{A symmetry for vanishing cosmological constant}

\author{Recai Erdem}
\email{recaierdem@iyte.edu.tr}
\affiliation{Department of Physics,
{\.{I}}zmir Institute of Technology \\ 
G{\"{u}}lbah{\c{c}}e K{\"{o}}y{\"{u}}, Urla, {\.{I}}zmir 35430, 
Turkey} 

\date{\today}

\begin{abstract} Two different realizations of a symmetry principle that 
impose a zero cosmological constant in an extra-dimensional set-up are 
studied. The symmetry is 
identified by  multiplication of  the metric by minus one. In the 
fist realization of the symmetry this is provided by a symmetry 
transformation that multiplies the coordinates by the imaginary number i. 
In the second realization this is accomplished by a symmetry 
transformation that multiplies the metric tensor by minus one. In both 
realizations of the symmetry the requirement of the invariance of the 
gravitational action under the symmetry selects out the dimensions given 
by $D=2(2n+1)$, $n=0,1,2....$ and forbids a bulk cosmological constant. 
Another 
attractive aspect of the symmetry is that it seems to be more promising 
for quantization when compared to the usual scale symmetry. 
The second realization of the symmetry principle is more 
attractive in that it is possible to make a possible brane cosmological 
constant zero in a simple way by using the same symmetry, and the symmetry 
may be identified by reflection symmetry in extra dimensions. 
\end{abstract}

\maketitle
The universe at cosmic scales may be described by a 
homogeneous and isotropic ideal fluid. The 00-component of the 
corresponding Einstein 
equations results in
\cite{Carroll1, Carroll2}
\begin{equation}
\frac{\mbox{\"{a}}}{\mbox{a}}\;=\; -\frac{4\pi\,G_N}{3}(\rho+3p) 
\label{t1}
\end{equation}
where a=a(t)$\,>\,0$ is the scale factor for the expansion of the universe 
related to the Hubble parameter $H$ by $H=\frac{\mbox{\.{a}}}{\mbox{a}}$, 
$G_N$ is the 
Newton's constant, $\rho$ is the energy density and $p$ is the pressure of 
the ideal fluid (modeling our universe at cosmic scales). Recent 
cosmological observations \cite{Astro} suggest that $\mbox{\"{a}}\,>\,0$ 
while 
the standard 
matter and radiation (e.g. stars and electromagnetic radiation) requires 
$\mbox{\"{a}}\,<\,0$. This combined with the amount of the standard matter 
and 
radiation requires a form of energy density with $p\simeq-\rho$, which, 
in turn, 
may 
be identified with vacuum energy density $\rho_v$ of value $\simeq 
(2,3\times 10^{-3}eV)^4$ \cite{PDG}. This is the 
most standard explanation for acceleration of the universe although 
there are alternative ways of explanation as well \cite{Copeland}. Vacuum 
energy density 
results in a stress-energy tensor that may be identified by a cosmological 
constant through the relation $\rho_v=\frac{\Lambda}{8\pi\,G_N}$. However 
the value of the theoretical contributions to vacuum energy density 
$\simeq (100 $Mev$)^4$ - $(10^{19}$Gev$)^4$ 
is extremely larger 
than its measured value 
$(\simeq 10^{-3}$eV$)^4$ \cite{Weinberg, Carroll2}. Most of the 
so-called cosmological constant problems (i.e. what is the source of the 
huge discrepancy between the theoretical and the observational values 
of $\Lambda$, why is $\Lambda$ so small?, why is $\Lambda$ not 
exactly equal to zero?) are variations of this fact 
In this 
talk I study only one of these cosmological constant problems, namely,  
why is $\Lambda$ so small?. In literature there are many different schemes 
that deal with this problem \cite{Weinberg,Nobbenhuis}; symmetries 
(i.e. supersymmetry, supergravity, superstrings, 
conformal symmetry, invariant length reversal symmetry), anthropic 
considerations, adjustment mechanisms, changing gravity, quantum 
cosmology, diluting through extra dimensions. In this study  
a symmetry principle in an 
extra dimensional set-up
is employed to make the cosmological constant zero. The accelerating 
expansion of the 
universe then may either be 
attributed to breaking of the 
symmetry by a small amount through the usual symmetry arguments or may be 
attributed to the alternative mechanisms of the acceleration 
\cite{Copeland}. 
I consider two different realizations of this symmetry. The fist 
realization 
employs a symmetry 
transformation that multiplies the coordinates by the 
imaginary number i \cite{Erdem1, Nobbenhuis, tHooft}. The second 
realization is implemented by signature reversal that multiplies the metric 
tensor by -1 \cite{Erdem2,Duff}. In both realizations the requirement of 
the (non-vanishing and the) invariance of the gravitational action 
restricts 
the number of space-time dimensions to $D=2(2n+1)$, $n=0,1,2,.....$ (or 
stating more precisely; to spaces that have a $2(2n+1)$ dimensional 
subspace whose metric being odd under the signature reversal and the 
metric of the remaining part of the space being even under  
signature reversal). The symmetry forbids a bulk cosmological 
constant in the allowed dimensions. A 
brane cosmological constant confined to the usual 4-dimensional space is 
forbidden by the symmetry because $D=4$ does not satisfy the rule 
$D=2(2n+1)$. However an effective 4-dimensional 
cosmological constant may be induced through the part of the curvature 
scalar, that depends only on the extra dimensions. In 
order to  forbid such a contribution to the cosmological constant one 
needs an extra mechanism in the first realization while in the second 
realization this can be achieved by putting the usual 4-dimensional space 
at the intersection of two $2(2n+1)$ dimensional spaces and then
imposing the same symmetry i.e. the signature reversal symmetry to both 
spaces as will be shown later in this talk. I also find that the form of 
the matter Lagrangian and the transformation rule for fields (other than 
gravitation) obtained under the requirement of the corresponding action 
functional are almost the same in both realizations. The transformation 
rules for the fields suggest that this symmetry is more promising for 
quantization than the usual scaling symmetry.

In this talk I consider a symmetry whose 
effect is to multiply the metric by minus one, that is,
\begin{equation}
ds^2\,=\,g_{AB}dx^A\,dx^B 
~\rightarrow 
~~-\,ds^2 \label{t2}
\end{equation}
The fist realization of this symmetry is through the transformation ( that 
multiplies the coordinates $x_A$ by i)
\cite{Erdem1, Nobbenhuis, tHooft}
\begin{equation}
 x_A \rightarrow i\, x_A~~~~~~~~~,~~~~~~A=0,1,2,......,D-1
\label{t3} 
\end{equation}
where $D$ is the dimension of the space. The requirement of the 
invariance of physics 
under the symmetry transformation (\ref{t3}) may be imposed in two ways; 
either through the requirement of the covariance of the Einstein field 
equations or by the requirement of the invariance of the 
corresponding action functional under the symmetry transformation given in
(\ref{t3}). The application 
of the first approach to the gravitation (i.e. the requirement of 
the covariance of the Einstein field equations under the transformation 
(\ref{t3})) results in the conclusion that the cosmological constant 
breaks the covariance of the Einstein equations and hence, is not 
allowed \cite{Nobbenhuis,tHooft}. This conclusion is independent of the 
number of dimensions of the 
space. Hence one may take the space be the usual 4-dimensional space. The 
second approach \cite{Erdem1} will be followed here and it  
leads to a restriction on the number of dimensions. In this 
approach we require the invariance of the gravitational action functional
\begin{equation}
S_R = \frac{1}{16\pi\,G}\int 
\sqrt{g} \,R\,d^Dx \label{t4}
\end{equation}
under (\ref{t3}). Here $g=(-1)^sdet(g)$, $s$ = 0 or 1 so that $\sqrt{g}$ 
gives a real number contribution to the 4-dimensional action after 
integration over extra dimensions.  One notes that 
\begin{equation}
R \rightarrow -R~~~~,~~~~~~ 
\sqrt{g} \,d^Dx \rightarrow
(\pm\,i)^D\,\sqrt{g} \,d^Dx ~~~~~~\mbox{as}~~~~~x_A\,\rightarrow\,i\,x_A
\label{t5} 
\end{equation}
So only the number of dimensions given by
\begin{equation}
D=2(2n+1)~~~,~~~~n=0,1,2,3,....
\label{t6}
\end{equation}
are allowed by 
the invariance of (\ref{t4}) under 
(\ref{t3}). A bulk cosmological constant is forbidden in the 
dimensions given in (\ref{t6}) since the corresponding action 
functional
\begin{equation}
S_C = \frac{1}{16\pi\,G}\int \sqrt{g} \,\Lambda \,d^Dx \label{t7}
\end{equation}
is not invariant under the symmetry transformation (\ref{t3}). However 
a possible contribution to the 4-dimensional cosmological constant 
through the part of curvature scalar, that depends only on the extra 
dimensions is not forbidden in this realization of the symmetry; 
one needs an additional symmetry to forbid it. Such a symmetry was 
employed in \cite{Erdem1} for a six dimensional metric. The 
second 
realization of the symmetry is more promising in this 
respect because the same symmetry may be also employed to forbid a 
possible 
contribution to the 4-dimensional cosmological constant through 
curvature scalar as we will see. 

The symmetry transformation for the second realization of this symmetry 
\cite{Erdem2,Duff}
is given by
\begin{equation}
g_{AB}\rightarrow -\,g_{AB} \label{t8}
\end{equation}
The curvature scalar $R$ and the invariant volume element 
$\sqrt{g}\,d^Dx$ transform exactly in the same way as in the first 
realization (\ref{t5}). So the second realization as well 
selects out the dimensions $D=2(2n+1)$ and forbids a bulk cosmological 
constant. In fact it is not essential that the dimension of space is 
$2(2n+1)$ to have the symmetry be applicable. The essential point is 
that the space should contain a subspace whose metric tensor transforms 
like (\ref{t8}) while the metric tensor of the remaining part of the 
space is invariant under the symmetry transformation. However such a 
choice would be ad hoc.

The main advantage of the realization is 
that the same symmetry 
may be used to forbid a possible contribution to the 4-dimensional 
cosmological constant, after integration over extra dimensions, through 
the piece of the curvature scalar that  
depends only on extra dimensions. To this end I take two 2(2n+1) 
dimensional 
spaces, say, one with 6 dimensions and the other with 10 dimensions, and 
the 
usual 4-dimensional space is taken at the intersection of these spaces. I 
require that the transformations of the 
metric tensors of each space under the signature reversal (\ref{t8}) leave 
the action invariant, both  under the separate and 
the simultaneous transformations on the two spaces. The requirement of 
the invariance of the action under the signature reversal of the metric 
tensors of each space separately, 
guarantees the absence of bulk cosmological constants 
while the requirement of the invariance of the action under the 
simultaneous signature reversals of the metrics of both spaces guarantees 
the absence of any possible contribution to the 4-dimensional 
cosmological constant through the part of the curvature scalar that 
depends only on the extra dimensions. This mechanism may be better 
seen through the following example. Consider the metric describing the 
union of two spaces of dimensions $D^\prime$ and $D^{\prime\prime}$ 
\begin{eqnarray}
ds^2&=&
\Omega_1(y)\Omega_2(z)
g_{\mu\nu}(x)\,dx^\mu dx^\nu\,+ 
\Omega_1(y)g_{ab}(w)\,dx^a dx^b\,+\,
\Omega_2(z)g_{cd}(w)\,dx^c dx^d
\label{t9} \\
\mbox{where}&&~~~
x=x^\mu~,~~y=x^a~,~~z=x^c~,~~w=y,z \nonumber \\
&&\Omega_1(y)=\Omega_1(y_1)= 
\,=\,\cos{k_1\,x_{5^\prime}}~~,~
~~~\mbox{and}~~~~ 
\Omega_2(z)=\Omega_2(z_1) = 
\,=\,\cos{k_2\,x_{6^{\prime\prime}}} \label{t10} \\
&&\mu,\nu=0,1,2,3~~;~~~~ 
a,b\,=\,4^\prime,5^\prime,....D^\prime-1~~;~~
~c,d\,=\,4^{\prime\prime},5^{\prime\prime},....D^{\prime\prime}-1 
\label{t11} \\
&&D^\prime=2(2n+1)~~,~~~D^{\prime\prime}=2(2m+1)~~~~~~n,m=1,2,3,.....
\label{t12} 
\end{eqnarray}
The overall dimension of the space is $D=2n+2m$.
We notice that $\Omega_1(y)$, $\Omega_2(z)$
are odd functions of $y$, 
$z$; respectively,  under the reflection about the point 
$k_{1(2)}\,x_{5^\prime(6^{\prime\prime})}=\frac{\pi}{2}$,
\begin{equation}
k_{1(2)}\,x_{5^\prime(6^{\prime\prime})}
\,
\rightarrow\,\pi-
k_{1(2)}\,x_{5^\prime(6^{\prime\prime})}
\label{t13}
\end{equation}
The application of (\ref{t13}) to one of the $2(2n+1)$ 
dimensional spaces induces a transformation of the metric tensor of that 
space exactly in the same way as given in (\ref{t8}). Hence the curvature 
scalar and the invariant volume element transform exactly in the same way 
as given in (\ref{t5}). On the other hand the 
application of (\ref{t13}) to both spaces simultaneously results in 
\begin{eqnarray} 
&&g_{\mu\nu}\rightarrow \,g_{\mu\nu} ~~~~~,~~~~~~
g_{ab}\rightarrow -\,g_{ab} ~~~~~,~~~~~~
g_{cd}\rightarrow -\,g_{cd} 
\label{t14}
\end{eqnarray}
where the indices $\mu$, $\nu$, $a$, $b$, $c$, $d$ run as given in 
(\ref{t11}). In fact (\ref{t14}) is not specific to this example and is 
the general transformation rule for the 
metric tensor of a space that consists of the union of two $2(2n+1)$ 
dimensional spaces where the there is signature reversal symmetry in each 
space. The 4-dimensional part of the curvature scalar 
$R_4=g^{\mu\nu}R_{\mu\nu}$,
the extra dimensional part of the curvature scalar, 
$R_e=g^{ab}R_{ab}$, and the invariant volume element $\sqrt{g}\,d^Dx$ 
transform under the simultaneous applications of the two 
transformations in (\ref{t14}) as 
\begin{eqnarray}
R_4 \rightarrow R_4~~~~,~~~~~~ R_e \rightarrow -R_e~~~~,~~~~~~ 
\sqrt{g} \,d^Dx \rightarrow \,\sqrt{g} \,d^Dx \label{t15} 
\end{eqnarray}
The transformation rule for the metric under (\ref{t14}) becomes
\begin{equation}
ds^2\,=\,g_{MN}dx^M\,dx^N =g_{\mu\nu}dx^\mu\,dx^\nu+g_{ab}dx^a\,dx^b 
\rightarrow\,
g_{\mu\nu}dx^\mu\,dx^\nu-g_{ab}dx^a\,dx^b \label{t16}
\end{equation}
It is evident from (\ref{t15}) that the contribution due to $R_e$ vanishes 
and the one due to $R_4$ survives so that we reach our goal of eliminating 
any contribution to the cosmological constant through the part of the 
curvature scalar that depends only on extra dimensions. In fact this 
conclusion is true for any metric in a space formed of two $2(2n+1)$ 
dimensional spaces so that the usual 4-dimensional space is at their 
intersection, and that obeys 
(\ref{t14}), and has 4-dimensional Poincar{\,{e}} invariance 
\cite{Rubakov} 
( since the 4-dimensional Poincar{\,{e}} invariance insures the metric 
tensors 
of the extra dimensions depend only on extra dimensions). For a more 
detailed discussion and calculations one may refer to 
\cite{Erdem2}). In other words the requirement of the invariance of the 
action functional under the application of the signature reversal on each 
$2(2n+1)$ dimensional space separately (through transformations of the 
form of (\ref{t8}) ) insures absence of bulk cosmological constant while 
the requirement of the invariance of the action functional under the 
application of signature reversal on both spaces simultaneously (through 
transformations of the form of (\ref{t15}) ) insures absence of any 
contribution to the 4-dimensional cosmological constant through the extra 
dimensional piece of the curvature scalar.

Transformation rules for fields (other than gravitation) under the 
symmetry is another important issue to be discussed because it is decisive 
in the invariance properties of n-point correlation functions of 
quantum field theory. We require the invariance of  
the action functional 
\begin{eqnarray}
S_L &=& \int 
\sqrt{g}
\,d^Dx\,{\cal L} \label{t17}
\end{eqnarray} 
where ${\cal L}$ denotes the Lagrangian
for the fields other than gravitation. This gives us transformation 
rule for the Lagrangian and this transformation rule, in turn, is used to 
determine the transformation rule for the fields by using the requirement 
of the invariance of the kinetic terms of the Lagrangian. In the first 
realization of the symmetry 
\begin{equation} \sqrt{g} \,d^Dx \rightarrow
(i)^D\,\sqrt{g} \,d^Dx ~~~~~~\mbox{so in 2(2n+1) dimensions this 
imposes}~~~~~~
{\cal L}\rightarrow\,-{\cal L} 
 \label{t18}
\end{equation}
In the second realization the transformation rule for ${\cal L}$ is the 
same as the first realization (\ref{t18}) when the  
transformation is applied to the metric tensor of each space 
separately while
the transformation rules for the invariant volume element and the 
Lagrangian when
the
transformation is applied to the metric tensors of both spaces 
simultaneously 
are
\begin{equation}
\sqrt{g} \,d^Dx \rightarrow
(i)^{4n}\,\sqrt{g} \,d^Dx
=\,\sqrt{g} \,d^Dx
 ~~~~~~\mbox{so}~~~~~~
{\cal L}\rightarrow\,{\cal L} 
 \label{t19}
\end{equation}
where the transformation rule
(\ref{t14} )) is used.
Hence in both realizations one obtains the same transformation for the 
scalars
\begin{equation}
\phi\,\rightarrow\,\pm\phi \label{t20}
\end{equation}
and the extra dimensional piece of the kinetic term drops out in the 
second realization if the space is taken as the union of two spaces 
where the usual 4-dimensional space lies at the intersection. The 
transformation rule for gauge fields in both realizations are
different. In the first realization only $U(1)$ gauge fields $B_A$ are 
allowed and transform as
\begin{equation}
F_{AB}\rightarrow 
\pm\,i\,F_{AB}~~~~~\mbox{and}~~~~B_A\rightarrow \,B_A \label{t21}
\end{equation}
while in the second realization all gauge fields are allowed and transform 
as
\begin{equation}
F_{AB}\rightarrow 
\,F_{AB}~~~~~\mbox{and}~~~~B_A\rightarrow \,B_A 
\label{t23} 
\end{equation}
In the first realization, fermions are allowed only on $2n+1$ dimensional 
spaces. 
The situation is essentially the same in the second realization as well. 
Fermions $\psi$ in both realizations transform (in $2n+1$ dimensions) as
\begin{equation}
\psi \rightarrow e^\alpha \psi \label{t24}
\end{equation}
where $\alpha$ is an overall constant phase. Moreover it was shown in 
\cite{Erdem2} that the part of the fermionic Lagrangian that depends only 
on extra dimensions do not pose a problem for cosmological constant 
problem in the second realization since it cancels out after 
integration over extra dimensions. 

Once the transformation properties of the fields are 
determined one can discuss the invariance properties of the n-point 
(correlation) functions of quantum field theory 
\begin{equation}
<0|\varphi_1(x_1)\varphi_2(x_2)........\varphi_n(x_n)|0> \label{t25}
\end{equation}
where $|0>$, and $\varphi_k(x_k)$ stand for the vacuum state, and a 
general field at the position $x_k$ in a $D$ dimensional space. It is 
evident from 
the equations (\ref{t20}-\ref{t23}) that the basic building blocks 
for 
Feynman diagrams, two-point functions (propagators) are always invariant 
and arbitrary n-point functions are invariant in most 
of the cases under the symmetry discussed here.

In this study I have reviewed a symmetry that insures a zero cosmological 
constant. The acceleration of the universe either may be attributed to 
breaking of the symmetry by a small amount \cite{Erdem1} or to one of the 
alternatives ways such as quintessence, phantom(ghost) etc. 
\cite{Copeland}. This symmetry has more attractive aspects compared to 
other symmetries employed. Supersymmetry and supergravity theories are 
broken by a large amount when compared to the upper bound on the 
observational value of the cosmological constant while there is no such 
problem for this symmetry. Conformal symmetry is also employed to make 
cosmological constant zero in literature. However quantization of 
conformal field theories is troublesome \cite{Coleman} while this symmetry 
seems to be more promising in this aspect as well, as we have seen. 
Usually signature reversal is accompanied with existence of ghost fields. 
However the signature reversal symmetry here may be identified by 
reflections in extra dimensions, that 
is, the so-called ghosts and the usual particles do not share the same 
position so they do not cause the usual troubles caused by the presence of 
ghosts (in addition to the usual particles). Therefore it does not suffer 
from the problems of E-parity models \cite{Linde,Kaplan,Moffat} that use a 
usual particle - ghost particle symmetry to eliminate the cosmological 
constant problem. I think these points 
make the 
signature reversal symmetry ( introduced in the context of extra 
dimensional models) an attractive possibility to be considered further.




\bibliographystyle{plain}

\end{document}